\documentclass[11pt,epsf]{article}

\usepackage{latexsym}
\usepackage{amssymb}
\usepackage{amsmath}
\usepackage{amsthm}
\usepackage{stmaryrd}
\usepackage{times}
\usepackage{cancel}
\usepackage{graphicx}

\def\beq{\begin{eqnarray}}
\def\eeq{\end{eqnarray}}
\def\bea{\begin{eqnarray*}}
\def\eea{\end{eqnarray*}}

\def\centeron#1#2{{\setbox0=\hbox{#1}\setbox1=\hbox{#2}\ifdim
\wd1>\wd0\kern.5\wd1\kern-.5\wd0\fi
\copy0\kern-.5\wd0\kern-.5\wd1\copy1\ifdim\wd0>\wd1
\kern.5\wd0\kern-.5\wd1\fi}}
\def\ltap{\;\centeron{\raise.35ex\hbox{$<$}}{\lower.65ex\hbox{$\sim$}}\;}
\def\gtap{\;\centeron{\raise.35ex\hbox{$>$}}{\lower.65ex\hbox{$\sim$}}\;}

\def\singleandthirdspaced{\baselineskip=\normalbaselineskip\multiply
    \baselineskip by 130\divide\baselineskip by 100}

\newcommand{\newc}{\newcommand}
\newc{\qbar}{{\overline q}}
\newc{\Kahler}{K\"ahler }
\newc{\deltaGS}{\delta_{\rm GS}}
\begin{document}
\begin{titlepage}
\begin{flushright}
{\large  SCIPP-2008/12\\
}
\end{flushright}

\vskip 1.2cm

\begin{center}

{\LARGE\bf Effects of $\theta$ on the deuteron binding energy and the triple-alpha process}

\vskip 1.4cm

{\large Lorenzo Ubaldi}
\\
\vskip 0.4cm
{\it Santa Cruz Institute for Particle Physics and
\\ Department of Physics, University of California,
     Santa Cruz CA 95064  } \\
%{\it $^b$Institute for Advanced Study, Princeton, New Jersey,  08540  }\\
%{\it $^c$Physics Department, Rutgers University, Piscataway, New Jersey}
\vskip 4pt

\vskip 1.5cm

\begin{abstract}
We study the effects that a non zero strong-CP-violating parameter $\theta$ would have on the deuteron and diproton binding energies and on the triple-alpha process. Both these systems exhibit fine tuning, so it is plausible that a small change in the nuclear force would produce catastrophic consequences. Such a nuclear force is here understood in the framework of an effective Lagrangian for pions and nucleons, and the strength of the interaction varies with $\theta$. We find that the effects are not too dramatic.
\end{abstract}

\end{center}

\vskip 1.0 cm

\end{titlepage}
\setcounter{footnote}{0} \setcounter{page}{2}
\setcounter{section}{0} \setcounter{subsection}{0}
\setcounter{subsubsection}{0}

%%%%%%%%%%%%%%%%%%%%%%%%%%%%%%%%%%%%%%%%%%%
%%%%%%%%%%%%%%%%%%%%%%%%%%%%
\singleandthirdspaced
 
\section{Introduction}
%The strong CP problem
The QCD Lagrangian includes a $\theta$-term which is usually written as
\beq \label{eq:Ltheta}
\mathcal{L}_{\theta} = -\frac{g^2 \theta}{32\pi^2} F \tilde{F}.
\eeq
For $\theta \neq 0$, this leads to CP-violation in the Strong Interaction. Measurements of the neutron electric dipole moment set the severe bound $|\theta|<10^{-10}$ \cite{Crewther:1979pi, Kawarabayashi:1980uh}. The lack of a satisfactory explanation, within the Standard Model, of why this should be the case is referred to as the Strong CP problem.

%The landscape
A few years ago, it has been realized that string theory may possess a large discretuum of stable and/or metastable ground states \cite{Susskind:2003kw, Banks:2003es}. In this ''landscape'', $\theta$ is probably a random variable and can take different values in different vacua. It has been argued that $\theta$ is not significantly constrained by anthropic considerations and that some natural explanation, such as an axion, is required \cite{Banks:2003es, Donoghue:2003vs}. In this paper we want to explore what would happen in a universe where $\theta$ is not zero. Would such a universe be very different from ours? This is similar in spirit to the work recently published by Jaffe et al. \cite{Jaffe:2008gd}. Varying the quark masses, they investigate which values satisfy the environmental constraint that the quark masses allow for for stable nuclei, making organic chemistry possible. In our case, instead, we fix the quark masses to the values
\beq
m_u = 4 \ \rm{MeV}, \qquad \qquad m_d = 7 \ \rm{MeV}
\eeq
and let $\theta$ vary. We study the effects that this would produce on the binding energies of two among the lightest nuclei, the deuteron and the diproton\footnote{The diproton does not actually exist as a bound state in nature, but the effect of $\theta$ could be such to bind it.}, and on the abundance of carbon and oxygen. Csoto, Oberhummer and Schlattl \cite{Csoto:2000iw, Oberhummer:2000zj} determined that the abundance of $^{12}C$ and $^{16}O$ is extremely sensitive to even small changes in the strength of the nucleon-nucleon force. The models they use to describe the N-N interaction in their study don't involve explicitly the angle $\theta$. They multiply the strength of the N-N force by a factor $p$, which they then vary from 0.996 to 1.004.
In our case, the tool for exploring the consequences of $\theta \neq 0$ on the systems just mentioned is provided by a sigma model, intended as an effective Lagrangian that describes the interactions between pions and nucleons. For nuclei like $^{12}C$ and $^{16}O$, the nuclear force is described by contact interactions, the strength of which depends on the pion mass, that in turn depends on $\theta$.

%Organization
The paper is organized as follows. First we write the sigma model Lagrangian and we derive formulae for the pion mass and the proton-neutron mass difference as functions of $\theta$. Then, we compute the correction to the binding energies of the deuteron and the diproton, and we study the consequences of varying $\theta$ on the triple-alpha process. We conclude with a few comments on the results.

\section{Lagrangian for Nucleon-Pion interactions}

\subsection {$\theta$-dependence in the quark mass matrix}

For the purposes of the following discussion, it is convenient to remove the term (\ref{eq:Ltheta}) from the Lagrangian by performing a rotation of the quark fields
\beq
u &\to& e^{i\phi_u} u \\
d &\to& e^{i\phi_d} d,
\eeq
such that
\beq \label{eq:phitheta}
\phi_u + \phi_d = \theta.
\eeq
This introduces an equivalent $\theta$ dependence in the quark mass matrix, that we write as $MU_0$, where
\beq
M = 
\begin{pmatrix}
m_u & 0 \\
0 & m_d
\end{pmatrix},
\qquad \qquad  U_0 =
\begin{pmatrix}
e^{i\phi_u} & 0 \\
0 & e^{i\phi_d}
\end{pmatrix}.
\eeq

\subsection{The sigma model}

The sigma model Lagrangian provides a framework for understanding the very low energy limit of QCD. We use the notation of the text by Srednicki \cite{Srednicki:2007qs}, and write our effective Lagrangian for pions and nucleons as
\begin{equation} \label{eq:Lagrangian}
\begin{split}
\mathcal{L} =& -\frac{1}{4}f^2_\pi {\rm Tr}[\partial_\mu U \partial^\mu U^\dag] + B_0 {\rm Tr} [(MU_0)U + (MU_0)^\dag U^\dag] \\
& + i\bar{N}\gamma_\mu\partial^\mu N - m_N \bar{N} (U^\dag P_L + UP_R)N \\
& -\frac{1}{2}(g_A - 1)i\bar{N}\gamma^\mu (U\partial_\mu U^\dag P_L + U^\dag \partial_\mu U P_R) N \\
& - c_1 \bar{N} ((MU_0)P_L + (MU_0)^\dag P_R) N - c_2 \bar{N}(U^\dag (MU_0)^\dag U^\dag P_L + U(MU_0)U P_R)N \\
& - c_3 {\rm Tr} ((MU_0)U+(MU_0)^\dag U^\dag)\bar{N}(U^\dag P_L + UP_R)N  \\
& - c_4{\rm Tr} ((MU_0)U-(MU_0)^\dag U^\dag)\bar{N}(U^\dag P_L - UP_R)N,
\end{split}
\end{equation}
where $U = e^{i\pi^a \tau^a /f_\pi}$, $\pi^a$ is the pion field, $\tau^a$ are the isospin matrices, $f_\pi = 92.4$ MeV is the pion decay constant, $N$ is the nucleon field, $P_L = \frac{1}{2}(1 - \gamma_5)$ and $P_R = \frac{1}{2}(1+\gamma_5)$ are the projection operators, $g_A = 1.27$ is the axial vector coupling, and $c_1, c_2, c_3, c_4$ are dimensionless constants. $B_0$ is a constant with dimension of [mass]$^3$ that can be determined from ratios of meson masses in $SU(3)$. Roughly speaking, $B_0 \sim \Lambda_{\rm QCD}^3$. In this paper we use $B_0 = 7.6 \times 10^6$ MeV$^3$. In the Lagrangian above we wrote all the possible terms that are invariant under $SU(2)_L \times SU(2)_R$, with the fields obeying the transformation rules
\beq
N_L \to L N_L, \qquad N_R \to R N_R, \qquad U \to LUR^\dag, \qquad (MU_0) \to R(MU_0)L^\dag,
\eeq
for $L$, $R$ in $SU(2)$.

\paragraph{The pion mass.}
We first obtain a formula for the mass of the pion as a function of $\theta$. We can start by writing
\beq \label{eq:expandU}
U = e^{i\pi^a \tau^a / f_\pi} = \cos \frac{|\vec{\pi}|}{f_\pi} + i \frac{\pi^a}{|\vec{\pi}|} \tau^a \sin \frac{|\vec{\pi}|}{f_\pi}.
\eeq
It will prove convenient also to adopt the following parametrization for the quark mass matrix
\beq \label{eq:parM}
MU_0 = A \mathbf{1}_2 + iB \mathbf{1}_2 + C \tau^3  + i D\tau^3.
\eeq
Using (\ref{eq:expandU}) and (\ref{eq:parM}), the potential $V$ in the Lagrangian (\ref{eq:Lagrangian}) reduces to
\beq \label{eq:tadpole}
V = -B_0 {\rm Tr} [(MU_0)U + (MU_0)^\dag U^\dag] = -B_0 \left[4A \cos\frac{|\vec{\pi}|}{f_\pi} - 4D \frac{\pi^3}{|\vec{\pi}|} \sin \frac{|\vec{\pi}|}{f_\pi}  \right].
\eeq
In order not to have a tadpole in $\pi^3$, we impose the condition $D=0$
\beq \label{eq:Dzero}
D = \frac{1}{2} {\rm Tr} \left[ \tau^3
\begin{pmatrix}
m_u \sin \phi_u & 0 \\
0 & m_d \sin \phi_d
\end{pmatrix}
\right] = \frac{1}{2}(m_u \sin \phi_u - m_d \sin \phi_d) = 0
\eeq
Solving (\ref{eq:phitheta}) and (\ref{eq:Dzero}) we find the useful relations
\beq \label{eq:cossin}
\sin \phi_u &=& \frac{m_d \sin \theta}{[m^2_u + m^2_d + 2m_um_d \cos \theta]^{1/2}} \\
\sin \phi_d &=& \frac{m_u \sin \theta}{[m^2_u + m^2_d + 2m_um_d \cos \theta]^{1/2}} \\
\cos \phi_u &=& \frac{m_u + m_d \cos \theta}{[m^2_u + m^2_d + 2m_um_d \cos \theta]^{1/2}} \\
\cos \phi_d &=& \frac{m_d + m_u \cos \theta}{[m^2_u + m^2_d + 2m_um_d \cos \theta]^{1/2}}. 
\eeq
Next we determine $A$
\beq
A = \frac{1}{2} {\rm Tr} 
\begin{pmatrix}
m_u \cos \phi_u & 0 \\
0 & m_d \cos \phi_d
\end{pmatrix}
 = \frac{1}{2}(m_u \cos \phi_u + m_d \cos \phi_d).
\eeq
We now have all the ingredients to get an expression for the pion mass. From eq. (\ref{eq:tadpole}), expanding $\cos\frac{|\vec{\pi}|}{f_\pi}$ to second order we find
\beq \label{eq:pionmass}
m^2_\pi = \frac{2 B_0}{f^2_\pi} [m^2_u + m^2_d + 2m_um_d \cos \theta]^{1/2}.
\eeq
Note that this is an even function of $\theta$, therefore CP conserving.
This formula generalizes and, for $\theta = 0$, reduces to the well known $m^2_\pi = \frac{2 B_0}{f^2_\pi} (m_u + m_d)$. Note that, varying $\theta$ from 0 to $\pi$, the pion mass decreases, and it attains a minimum at $\theta = \pi$.

All this was done in $SU(2)$. One could be more ambitious and try to find a formula for the pion mass in $SU(3)$. In that case, the analysis is carried out in the same way. Requiring the absence of tadpoles translates into two conditions
\beq \label{eq:msin}
m_u \sin \phi_u = m_d \sin \phi_d = m_s \sin \phi_s,
\eeq
and Eq. (\ref{eq:phitheta}) is modified to
\beq \label{eq:phases}
\phi_u + \phi_d + \phi_s = \theta.
\eeq
Now (\ref{eq:msin}) and (\ref{eq:phases}) cannot be solved analytically, but if we make the reasonable approximation $m_u,m_d \ll m_s$, they reduce to
\beq
\phi_u +\phi_d &=& \theta \\
\phi_s &=& 0 \\
m_u \sin \phi_u &=& m_d \sin \phi_d,
\eeq
which can be solved, leading to the same solution we found previously. The pion mass then turns out be the same as in the $SU(2)$ case.

\paragraph{The nucleons.}
Let's now examine the part of the Lagrangian involving the nucleons. First we can rewrite it in a more convenient way, using the following field redefinition \footnote{This is the same field redifinition that the reader finds in \cite{Srednicki:2007qs}}
\beq
N = (u_0uP_L + u_0^\dag u^\dag P_R) \mathcal{N},
\eeq
where $u_o^2 = U_0$ and $u^2 = U$. The last five lines in (\ref{eq:Lagrangian}) become
\begin{equation} \label{eq:nucleons}
\begin{split}
& i\bar{\mathcal{N}} \gamma^\mu\partial_\mu \mathcal{N} -m_N \bar{\mathcal{N}} \mathcal{N} + \bar{\mathcal{N}} \gamma^\mu v_\mu \mathcal{N} -g_A\bar{\mathcal{N}} \gamma^\mu \gamma_5 a_\mu \mathcal{N} \\
& - \frac{1}{2} c_+ \bar{\mathcal{N}} (u(MU_0)u + u^\dag (MU_0)^\dag u^\dag)\mathcal{N} \\
& +\frac{1}{2}c_- \bar{\mathcal{N}} (u(MU_0)u - u^\dag (MU_0)^\dag u^\dag)\gamma_5\mathcal{N} \\
& -c_3{\rm Tr}[(MU_0)U + (MU_0)^\dag U^\dag]\bar{\mathcal{N}}\mathcal{N} +c_4{\rm Tr}[(MU_0)U - (MU_0)^\dag U^\dag]\bar{\mathcal{N}} \gamma_5\mathcal{N},
\end{split}
\end{equation}
where $v_\mu = \frac{i}{2} [u^\dag(\partial_\mu u) + u(\partial_\mu u^\dag)]$, $a_\mu =\frac{i}{2} [u^\dag(\partial_\mu u) - u(\partial_\mu u^\dag)]$, and $c_{\pm} = c_1 \pm c_2$. This is not yet particularly illuminating. With some more algebra, we can write, to lowest order, the corrections to the nucleon mass
\begin{equation} \label{eq:Lmass}
\begin{split}
\mathcal{L}_{mass} =& -\frac{1}{2} (c_++4c_3) [m^2_u +m^2_d + 2 m_u m_d \cos \theta]^{1/2}\bar{\mathcal{N}} \mathcal{N} \\
& +i(c_-+4c_4) \frac{m_um_d \sin \theta}{[m^2_u +m^2_d + 2 m_u m_d \cos \theta]^{1/2}} \bar{\mathcal{N}}\gamma_5 \mathcal{N} \\
& -\frac{1}{2} c_+ \frac{m_u^2-m_d^2}{[m^2_u +m^2_d + 2 m_u m_d \cos \theta]^{1/2}} \bar{\mathcal{N}}\tau^3 \mathcal{N}
\end{split}
\end{equation}
and the nucleon-pion interactions
\begin{equation} \label{eq:Lint}
\begin{split}
\mathcal{L}_{int} = & -ig_{\pi NN}\pi^a \bar{\mathcal{N}} \tau^a \gamma_5 \mathcal{N} + \frac{i}{2} c_- [m^2_u +m^2_d + 2 m_u m_d \cos \theta]^{1/2} \bar{\mathcal{N}} \frac{\pi^a \tau^a}{f_\pi} \gamma_5 \mathcal{N} \\
& + c_+ \frac{m_u m_d \sin \theta}{[m^2_u +m^2_d + 2 m_u m_d \cos \theta]^{1/2}} \bar{\mathcal{N}} \frac{\pi^a \tau^a}{f_\pi} \mathcal{N} \\
& + \frac{i}{2 f_\pi} (c_-+4c_4) \frac{m^2_u-m^2_d}{[m^2_u +m^2_d + 2 m_u m_d \cos \theta]^{1/2}} \pi^3 \bar{\mathcal{N}} \gamma_5 \mathcal{N}.
\end{split}
\end{equation}
From (\ref{eq:Lmass}) we get the proton-neutron mass difference
\beq \label{eq:pndiff}
m_n - m_p = c_+ \frac{m^2_d - m^2_u}{[m^2_u +m^2_d + 2 m_u m_d \cos \theta]^{1/2}}.
\eeq
Note that, varying $\theta$ from 0 to $\pi$, $m_n - m_p$ increases. It reaches the maximum value $c_+ (m_d+m_u)$ at $\theta = \pi$.

\paragraph{Estimation of the constants.}
The constants $c_+, c_-, c_3, c_4$ can in principle be related to quantities measured in experiments. Since in our world $\theta$ is smaller than $10^{-10}$ (see \textit{e.g.} \cite{Crewther:1979pi}), we define these quantities to be measured at $\theta = 0$. Note that, with this definition, we do not learn anything about $c_-$ and $c_4$ from the nucleon mass, since the second term in (\ref{eq:Lmass}) vanishes at $\theta=0$. It would be good, for the sake of completeness if nothing else, if we could determine all the constants, but this task is not so easy and, for the calculation that we will perform in the next section, only $c_+$ contributes substantially.

The value of $c_+$ can be estimated from the measured proton-neutron mass difference ($\sim 1.3$ MeV at $\theta = 0$). Taking into account also the electromagnetic contribution $\epsilon_{\rm EM} \sim 0.5$ MeV we have
\beq
(m_n - m_p)_{\rm measured} = c_+ (m_d - m_u) - \epsilon_{\rm EM},
\eeq
yielding $c_+ = 0.6$. This estimation is crude, because the second contribution on the right hand side of the above equation is of the same order as the first one. It is more accurate to look at the mass splitting $M_\Xi - M_N$ in the baryon octet, as pointed out in \cite{Crewther:1979pi}. That yields $c_+ = 2.5$. This is the value that we are going to use in the next section.
The constant $c_-$ deserves some comments. If we look at the first line of (\ref{eq:Lint}), it appears that $\frac{c_-}{2f_\pi} [m^2_u +m^2_d + 2 m_u m_d \cos \theta]^{1/2}$ can be considered as some kind of correction to $g_{\pi NN}$. At $\theta=0$, one could interpret the measured value of $g_{\pi NN}$ as including such a correction, but that would not tell us anything about $c_-$. In other words, one could trade $c_-$ for a new constant, say $g'_{\pi NN} = g_{\pi NN} + \frac{c_-}{2f_\pi} (m_u +m_d)$. For $\theta \neq 0$, though, one wants to keep the contribution coming from $c_-$ separate from $g_{\pi NN}$ and deal with the fact that there seems to be no obvious physical quantity from which this constant can be estimated. From the construction of the Lagrangian, it makes sense to believe that $c_-$ should be of the same order as $c_+$, namely of order unity, because they are both linear combinations of $c_1$ and $c_2$ that appear in eq.(\ref{eq:Lagrangian}), but there is no proof of this. On the other hand, a value as big as 10 would be disturbing because it would cancel the suppression $\frac{[m^2_u +m^2_d + 2 m_u m_d \cos \theta]^{1/2}}{f_\pi} \sim \frac{1}{10}$. As already stated, we will not need $c_-$ for our calculation. We actually need to make this statement more precise: we can forget about the exact value of $c_-$ as long as it is not much greater than one. The reason for this will be discussed in the next section. 

% The measured value of $g_{\pi NN}$ could be interpreted as including such a correction (with $\theta=0$), but that would not tell us anything about $c_-$. We are free to define, instead, that the measured value of $g_{\pi NN}$ does not include any contribution from $c_-$. In this case, we can take $c_-$ to be of order 1, at the most, so that we are safe and do not introduce an appreciable correction to $g_{\pi NN}$. For the calculations in the next section, thus, we will use $c_- = 1$.  

Eq.(\ref{eq:pionmass}) and Eq.(\ref{eq:pndiff}) are the main results of this section. They make the $\theta$-dependence of the pion mass and the proton-neutron mass difference explicit and, since these quantities play key roles in determining nuclear properties, they can be used to explore the consequences of a non-zero strong-CP-violating parameter in nuclear physics.

\section{Effects of $\theta$ in nuclear physics}

We don't have yet a complete picture to explain nuclear physics in terms of effective field theories, but enough progress has been made to allow us to investigate, at least qualitatively, the effects that $\theta \neq 0$ would have in nuclear physics. In this section we focus our attention on:
\begin{itemize}
\item \textit{Two-nucleon systems}, namely the deuteron and the diproton. The former has a binding energy which is relatively small (2.2 MeV); the latter doesn't exist as a bound state in nature, but we know that it fails to bind by only $\sim 70$ keV. In principle, one expects that the $\theta$-dependent nucleon-pion interactions in (\ref{eq:nucleons}) could give corrections to these energies that might be big enough to unbind the deuteron or to bind the diproton. If either one of these possibilities were realized, the consequences would be dramatic. For instance, if the diproton were bound, all the hydrogen in the Universe would have been burnt to He$^2$ during the early stages of the Big Bang and no hydrogen compounds or stable stars would exist today. Likewise, an unbound deuteron would significantly change the chain of nucleosynthesis that leads to heavier elements \cite{Barrow}. Another reason for studying these two-nucleon systems is that they are simple, and we have a good control over the calculation. Other authors have studied the dependence of the deuteron binding energy on variations of other parameters, such as the coupling constant \cite{Dent:2001ga}, or the quark masses \cite{Golowich:2008wn}; 
\item \textit{The triple-alpha process}, which is responsible for the production of carbon in stars. The observed abundance of carbon and oxygen results from a peculiar position of various nuclear energy levels, and it is very sensitive to even small shifts of such levels. It is hard to relate the spacing between excited states of a nucleus to first principles, but, with some assumptions, we can qualitatively study how variations of $\theta$ affect the triple-alpha process. 
\end{itemize}

\subsection{Two-nucleon systems}

\paragraph{The deuteron.}

The deuteron exists as a bound state only in an isospin singlet and spin triplet configuration, and its binding energy is rather small ($E = -2.22$ MeV). Attempts to derive the nuclear potential starting from a Chiral Lagrangian show that the deuteron binding is predominantly a consequence of two-pion and three-pion exchanges. The two-pion can be modeled by $\sigma (600)$ exchange, which gives an attractive medium-range contribution, whereas the three-pion corresponds to an $\omega (783)$ exchange, that is short-range and repulsive. The one-pion exchange is responsible for the long-range contibution. 

For the purpose of our study here, however, we can content ourselves with a much simpler form for the potential, a three-dimensional square well
 \begin{displaymath}
    V(r) = \left\{ \begin{array}{ll}
        -V_0 & r<R \\
        0 & r>R
      \end{array} \right.;
    % \end{displaymath}
    % \begin{displaymath}
    \qquad
    \psi(r) = \left\{ \begin{array}{ll}
        A \frac{\sin kr}{r} & r<R \\
        B \frac{e^{-\rho r}}{r} & r>R
      \end{array} \right.
  \end{displaymath}
  with parameters chosen to fit the experimental measurements: $V_0 = 41$ MeV, $R = 8.62 \times 10^{-3}$ MeV$^{-1}$, $k = 212$ MeV, $\rho = 46.4$ MeV, $A = 2.31$ MeV$^{1/2}$, $B = 1.44 A$.

\begin{figure}[htp]
\centering
\includegraphics[width=120mm]{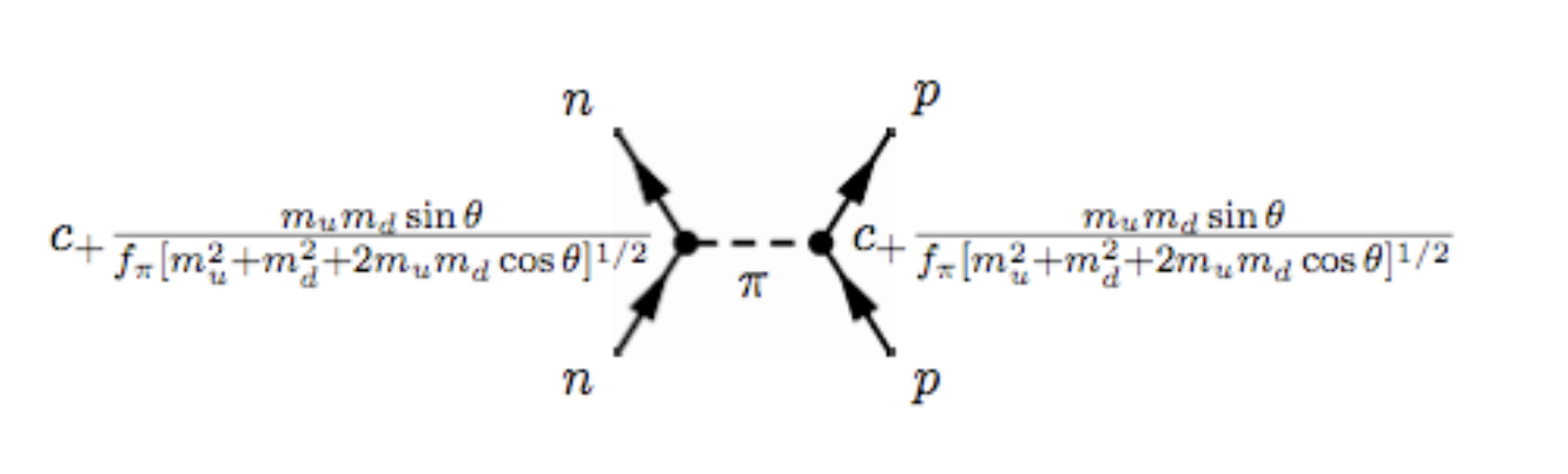}
\caption{\small \sl Feynman diagram}
  \label{fig:graphs}
\end{figure}

We want to compute the first-order corrections to the potential that we get from the theta-dependent terms in the Lagrangian, and see how significant they are. The interaction terms, that we need to look at, are listed in (\ref{eq:Lint}). A couple of comments are in due order:
\begin{itemize}
\item all the terms in (\ref{eq:Lint}), except for the first one, are suppressed by $\frac{m_q}{f_\pi}$, where $m_q$ stands for the quark mass and is, roughly speaking, a few MeV;
\item the terms containing a $\gamma_5$ get an extra spin-suppression that goes as $\frac{m_\pi}{2m_N}$ at each nucleon-nucleon-pion vertex. Note that $\frac{m_\pi}{2m_N}$ is of the same order as $\frac{m_q}{f_\pi}$.
\end{itemize}
Thus, a one-pion exchange diagram with $g_{\pi NN}$ at one vertex and $c_- \frac{m_q}{f_\pi}$ at the other vertex is suppressed with respect to a diagram with $c_+ \frac{m_q}{f_\pi}$ at both vertices, as long as $c_-$ \textit{is at the most of order 1}. 
To lowest order, then, we only need to evaluate the diagram shown in Figure \ref{fig:graphs}. 
In the non-relativistic limit it gives
\beq \label{eq:evdia}
 + ic_+^2 \frac{m^2_um^2_d \sin^2 \theta}{f^2_\pi[m^2_u +m^2_d + 2 m_u m_d \cos \theta]}  \frac{\vec{\tau}_n \cdot \vec{\tau}_p}{\mathbf{q}^2+m_\pi^2} 
\eeq
where $\mathbf{q}$ is the three-momentum of the exchanged pion. Using $\vec{\tau}_n \cdot \vec{\tau}_p = -3$ for the isosinglet, and Fourier transforming to position space, we find the following correction to the potential
\begin{equation} \label{eq:correction}
V_1(r, \theta) = \frac{3}{4\pi} \frac{c_+^2}{f^2_\pi} \frac{m^2_um^2_d \sin^2 \theta}{[m^2_u +m^2_d + 2 m_u m_d \cos \theta]} \frac{e^{-m_\pi r}}{r}.
\end{equation}
This is repulsive for all values of $\theta$. We can now use this result to compute the shift in the deuteron binding energy using first order perturbation theory
\beq
\Delta(\theta) = \langle \psi(r)|V_1(r, \theta)|\psi(r) \rangle .
\eeq
\begin{figure}[htp]
\centering
\includegraphics{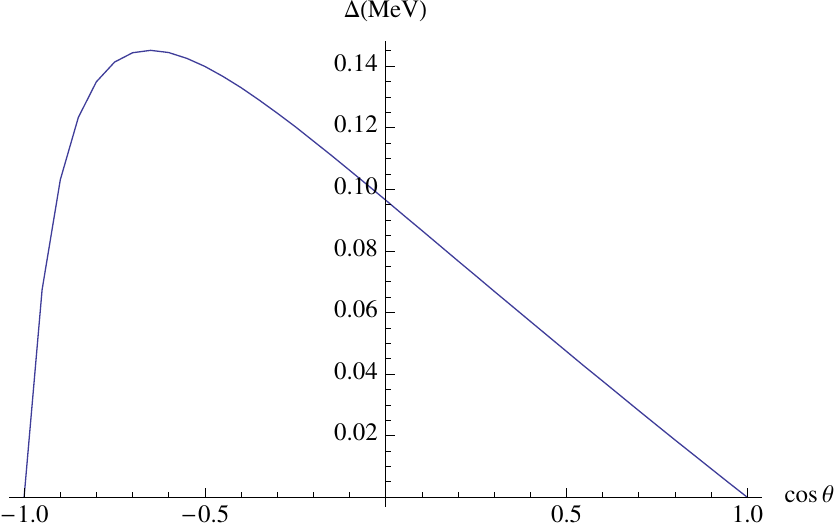}
\caption{\small \sl Shift in the deuteron binding energy as a function of $\cos \theta$} \label{fig:deuteron}
\end{figure}
The energy shift is plotted as a function of $\theta$ in Figure \ref{fig:deuteron}. At $\cos \theta \simeq -0.65$, we read from the plot that the binding energy is reduced by $\Delta \simeq 0.15$ MeV. This is a small number compared to 2.22 MeV, but might still have an appreciable effect on the early stages of Big Bang Nucleosynthesis, since the reaction rates depend exponentially on the deuteron binding energy.

\begin{figure}[htp]
\centering
\includegraphics{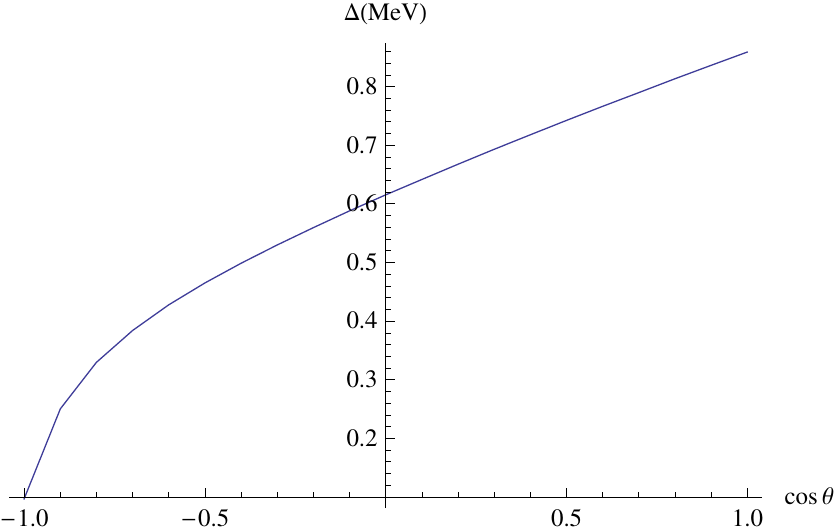}
\caption{\small \sl Shift in the deuteron binding energy with $c_- = 10$} \label{fig:c-}
\end{figure}

Before studying the diproton, let's see what would happen if $c_-$ was 10 instead of order 1. In this case, we would have $c_- \frac{m_q}{f_\pi} \sim 1$, and the diagram with $g_{\pi NN}$ at one vertex and $c_- \frac{m_q}{f_\pi}$ at the other vertex would not be suppressed anymore with respect to the one in Figure \ref{fig:graphs}. Including its contribution we would come to a qualitatively different conclusion as shown in Figure \ref{fig:c-}: the maximum value of the energy shift would be at $\theta = 0$.

% but we have reasons to believe that such an effect would be appreciable in the early stages of Big Bang Nucleosynthesis, since the reaction rates depend exponentially on the deuteron binding energy.

\paragraph{The diproton.}

\begin{figure}[htp]
\centering
\includegraphics{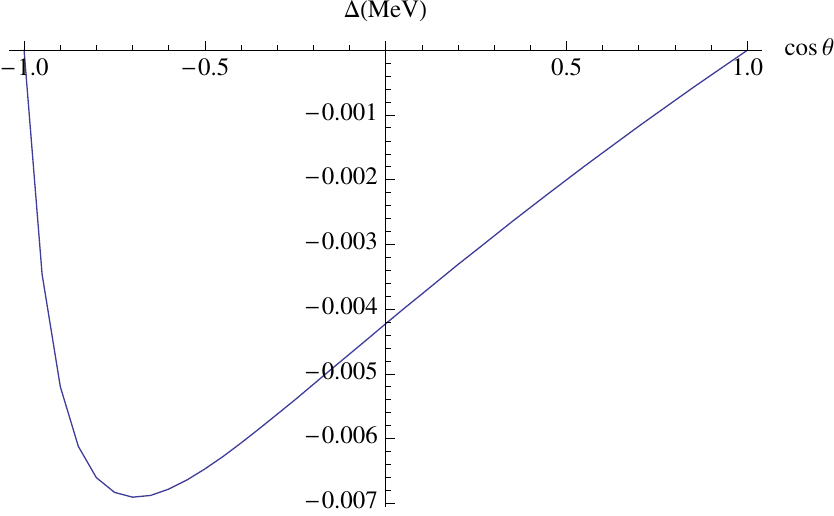}
\caption{\small \sl Shift in the diproton binding energy as a function of $\cos \theta$} \label{fig:diproton}
\end{figure}

The diproton almost exists as a bound state, so it is conceivable that the correction to the potential, that we get by calculating the diagram analogous to the one in Figure \ref{fig:graphs} (just replacing the neutron with a proton), might be significant enough to bind this system. We will proceed along the same line as for the deuteron. Here we adopt again a three-dimensional square well potential with the following parameters: $V_0 = 14$ MeV, $R = 13.1 \times 10^{-3}$ MeV$^{-1}$, $k = 114$ MeV, $\rho = 8.2$ MeV, $A = 1.09$ MeV$^{1/2}$, $B = 1.11 A$. With this choice of parameters, the diproton fails to be bound by an energy $E=72$ keV. The evaluation of the Feynman diagram in the non-relativistic limit gives (\ref{eq:evdia}), with $\vec{\tau}_n \cdot \vec{\tau}_p$ replaced by $\vec{\tau}_p \cdot \vec{\tau}_p$. If the diproton were bound, it would be in an isosinglet state, in which case $\vec{\tau}_p \cdot \vec{\tau}_p = +1$. The correction to the potential is then the following
\begin{equation}
V_1(r, \theta) = -\frac{1}{4\pi} \frac{c_+^2}{f^2_\pi} \frac{m^2_um^2_d \sin^2 \theta}{[m^2_u +m^2_d + 2 m_u m_d \cos \theta]} \frac{e^{-m_\pi r}}{r},
\end{equation}
and is attractive. The energy shift $\Delta(\theta) = \langle \psi(r)|V_1(r, \theta)|\psi(r) \rangle$ is plotted in Figure \ref{fig:diproton}.
At $\cos \theta \simeq -0.69$, $\Delta$ attains the minimum value of $\sim$ -7 keV. This represents a 10\% correction to the energy, which is not enough to bind the diproton, but might still have consequences on the early stages of the Nucleosynthesis chain.

To summarize, the effects of the angle $\theta$ on the binding energies of the deuteron and the diproton are at most 10\% corrections (6-7\% for the deuteron), which could be significant enough to affect the early stages of the Big Bang Nucleosynthesis. So large values of $\theta$ might have appreciable consequences.

\subsection{The triple-alpha process}

\begin{figure}[htp]
\centering
\includegraphics{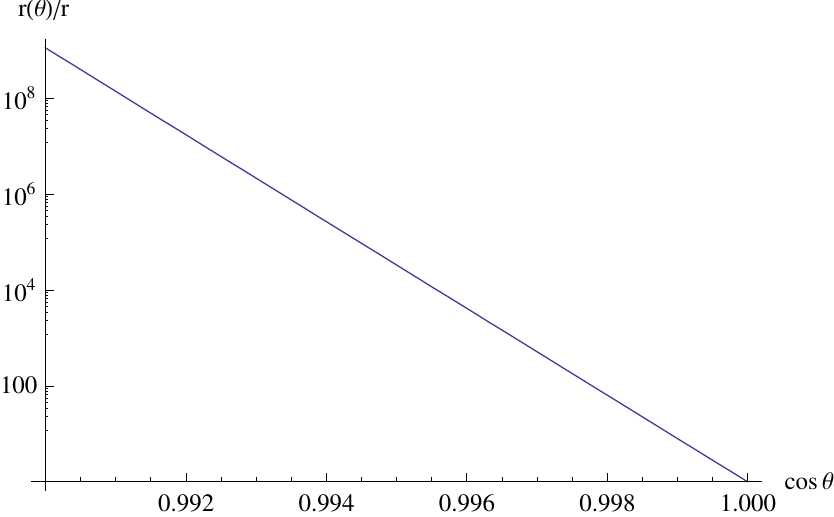}
\caption{\small \sl Reaction rate for the triple-alpha process as a function of $\cos \theta$.} \label{fig:rratetheta}
\end{figure}

\begin{figure}[htp]
\centering
\includegraphics{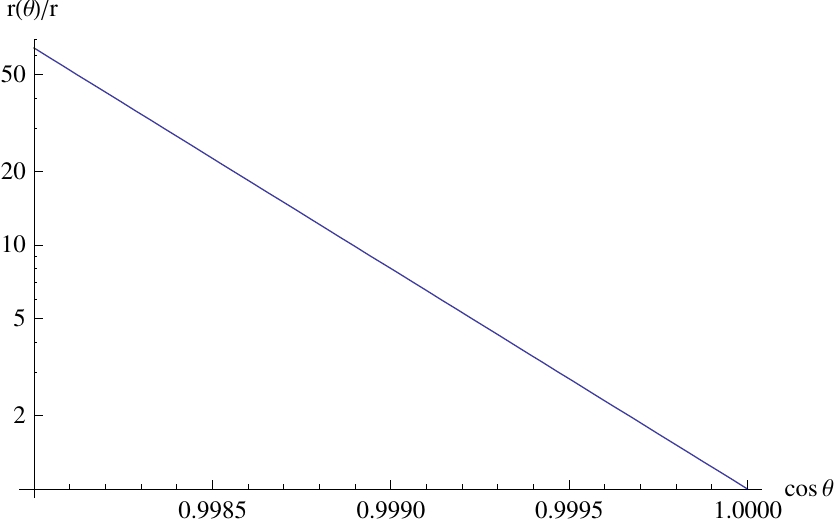}
\caption{\small \sl Reaction rate for the triple-alpha process as a function of $\cos \theta$. Here we plot a narrower range to see the detail for smaller values of $\theta$.} \label{fig:rratethetanarrow}
\end{figure}

The production of carbon in stars results from the reaction $3\alpha \leftrightarrow \alpha + ^8Be \leftrightarrow ^{12}C^{**}$. The $^8Be$ nucleus, in the second step, is unbound, but it lives long enough to allow for the possibility of capturing another alpha particle to form $^{12}C$. However, to produce the observed abundance of Carbon, this second reaction must be resonant. The $0_2^+$ state of $^{12}C$, lying at 380 keV, relative to the $3\alpha$ threshold (and 7654 keV above the $^{12}C$ ground state), provides such a resonance. The reaction rate for the triple-alpha process goes as \cite{Rolfs:88}
\beq 
r \sim \frac{\Gamma_\alpha \Gamma_{\rm rad}}{\Gamma} \exp \left(\frac{-Q_{3\alpha}}{T}\right)
\eeq
where
\beq \label{eq:Q}
Q_{3\alpha} = M_{^{12}C^{**}}-3M_\alpha ,
\eeq
$\Gamma_\alpha$ is the alpha particle width, $\Gamma_{\rm rad}=\Gamma_\gamma + \Gamma_{\rm pair}$ is the sum of electromagnetic decay widths to the $^{12}C$ ground state via gamma-ray emission or via electron-positron pair emission and $\Gamma = \Gamma_\alpha + \Gamma_{\rm rad}$. The following approximations hold: (i) $\Gamma_\alpha \gg \Gamma_{\rm rad}$ and (ii) $\Gamma_{\rm rad} \simeq \Gamma_\gamma$, so that $\frac{\Gamma_\alpha \Gamma_{\rm rad}}{\Gamma}\simeq \Gamma_\gamma$ and we can write
\beq \label{eq:rrate}
r \sim \Gamma_\gamma \exp \left(\frac{-Q_{3\alpha}}{T}\right).
\eeq
The measured values that enter the above equation are $Q = 380$ keV, $T \simeq 10$ keV and $\Gamma_\gamma \simeq 3.6$ meV. Let's take these to be our values at $\theta = 0$ and let's now see what would happen if $\theta$ were not zero. For simplicity, we make the assumption that the energy of the excited state $^{12}C^{**}$ with respect to the ground state $^{12}C$ does not vary with $\theta$. This assumption is probably unrealistic, but we use it to get a feeling for the various possibilities. It follows that $\Gamma_\gamma$ is nearly constant as well. But a small variation of $Q_{3\alpha}$ can have significant effects, because it appears in the exponential. We have
\beq
M_{^{12}C^{**}} = 6m_p + 6m_n + BE_{C} + 7.654, \qquad \qquad \quad M_\alpha = 2m_p + 2m_n + BE_{\alpha},
\eeq 
where $BE$ is the binding energy (negative) and everything is measured in MeV. Thus
\beq \label{eq:Q1}
Q_{3\alpha} = BE_{C} - 3BE_{\alpha} + 7.654.
\eeq
Following the work done by Furnstahl and Serot \cite{Furnstahl:1999rm}, and by Donoghue and Damour \cite{Damour:2007uv}, we can parametrize the binding energy per nucleon $BE/A$ as \cite{Damour:2007uv}
\beq \label{eq:binding}
\frac{BE}{A} = -(120 - \frac{97}{A^{1/3}}) \eta_S + (67 - \frac{57}{A^{1/3}})\eta_V + {\rm residual} \ {\rm terms}.
\eeq
This formula comes from considering the nuclear force as due to contact interactions. For all but the lightest nuclei, the key aspect of binding comes from a spin singlet and isospin singlet central potential, for which one can write a scalar and a vector contribution
\beq
H_{\rm{contact}} = G_S(\bar N N)(\bar N N) + G_V(\bar N \gamma_\mu N)(\bar N \gamma^\mu N),
\eeq
where $G_S$ is negative (i.e. attractive) and $G_V$ is positive (i.e. repulsive). In the traditional meson exchange models, the scalar component corresponds to the exchange of the $\sigma(600)$ meson and the vector component to the exchange of the $\omega(783)$ meson. We define $\eta_S$ and $\eta_V$, that appear in Eq. (\ref{eq:binding}), as\footnote{Our $G_{S,V}(\theta=0)$ is the same as what Damour and Donoghue \cite{Damour:2007uv} call $G_{S,V}|_{\rm {physical}}$.}
\beq
\eta_S &\equiv &\frac {G_S(\theta)}{G_S(\theta=0)} \\
\eta_V &\equiv &\frac {G_V(\theta)}{G_V(\theta=0)}.
\eeq
The scalar channel is the only portion of the central force that receives large effects from low energy. The sensivity of the vector channel to $m^2_\pi$ leads to sub-leading corrections compared to the effects linked to the $m^2_\pi$ sensitivity of the scalar channel (the reader should refer to \cite{Donoghue:2006du} for the details). For this reason, we will take $\eta_V = 1$ for our discussion and focus on the dominant scalar-channel effects. We parametrize $\eta_S$ from Figure 2 in \cite{Damour:2007uv}
\beq
\eta_S = -0.4\frac{m^2_\pi(\theta)}{m^2_{\rm phys}} + 1.4,
\eeq
where $m^2_{\rm phys} = m^2_\pi(\theta = 0)$ is the physical mass of the pion. The residual terms in Eq. (\ref{eq:binding}), which we assume not to depend on $\theta$, take care of all the other contributions that are not encoded by $\eta_S$ or $\eta_V$, such as the Coulomb repulsion, for example, and can be adjusted to get the measured $BE/A$ for each element at $\eta_S = \eta_V = 1$. For $^{12}C$ we have
\beq
\left( \frac{BE}{A} \right)_C = -77.631 \eta_S + 69.965,
\eeq
for $^4He$
\beq
\left( \frac{BE}{A} \right)_\alpha = -58.894 \eta_S + 51.834.
\eeq
Thus, we can write $Q_{3\alpha}$ as a function of $\theta$
\beq \label{eq:Q2}
Q_{3\alpha}(\theta) &=& 12\left( \frac{BE}{A} \right)_C - 12 \left( \frac{BE}{A} \right)_\alpha + 7.654\\
&=& 89.938 \left(\frac{[m^2_u + m^2_d + 2m_um_d \cos \theta]^{1/2}}{m_u+m_d} \right) - 89.556.
\eeq
For the resonant reaction to occur, $Q_{3\alpha}$ must be a positive quantity, which is equivalent to require that the excited state $^{12}C^{**}$ be above threshold. The condition $Q_{3\alpha}>0$ translates into the constraint
\beq
\cos \theta > 0.982 \qquad \qquad (\theta < 11^\circ).
\eeq
We can plot $r(\theta)/r$ vs $\cos \theta$, where $r(\theta)$ is the reaction rate (\ref{eq:rrate}) as a function of $\theta$ and $r \equiv r(\theta=0)$. The result is shown in Figure \ref{fig:rratetheta} and Figure \ref{fig:rratethetanarrow}.%  We report some values of $r(\theta)/r$ to give an idea
% \bea
% & \theta  \quad \qquad & r(\theta)/r \\
% & 2.56^\circ \quad \qquad &  8.02 \\
% & 3.62^\circ \quad \qquad &  64.3 \\
% & 4.44^\circ \quad \qquad &  516 \\
% & 5.13^\circ \quad \qquad &  4.14 \times 10^3 \\
% & 6.28^\circ \quad \qquad &  2.67 \times 10^5 \\
% & 7.25^\circ \quad \qquad &  1.73 \times 10^7.
% \eea

We see that for $\theta \neq 0$ the reaction rate increases dramatically. This would lead to a greater abundance of carbon in the universe.

Carbon is then involved in the reaction $^{12}C + \alpha \to ^{16}O$ to produce oxigen. In a world where $\theta = 0$, there is no energy level in $^{16}O$ to allow for this last reaction to be resonant, and that is why a substantial amount of $^{12}C$ survives. The closest level that could give a resonance is 2.42 MeV above the $^{12}C + \alpha$ threshold, too high to be resonant. There are two levels that are just sub-threshold, though, one at -45 keV, the other at -245 keV. It is conceivable that in our framework, when we vary $\theta$, we shift these levels enough to allow for a resonant reaction that would burn most of the carbon to form oxygen. Let's check if this happens.

\begin{figure}[htp]
\centering
\includegraphics{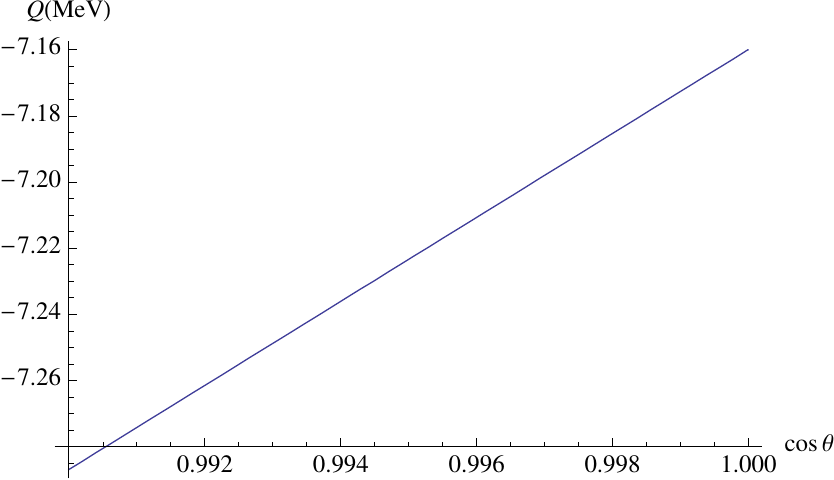}
\caption{\small \sl $Q$-value as a function of $\theta$ for the reaction $^{12}C + \alpha \to ^{16}O$} \label{fig:Oxygen}
\end{figure}

We assume again that the energy of the excited states is fixed with respect to the ground state of $^{16}O$, and we consider the $Q$-value for the reaction $^{12}C + \alpha \to ^{16}O$
\beq
Q(\theta) = M_O - M_C - M_\alpha = 16\left( \frac{BE}{A} \right)_O - 12\left( \frac{BE}{A} \right)_C - 4 \left( \frac{BE}{A} \right)_\alpha
\eeq
where 
\beq
\left( \frac{BE}{A} \right)_O = -81.505 \eta_S + 73.543.
\eeq
At $\theta = 0$ we get the measured result $Q=-7.16$ MeV. In Figure \ref{fig:Oxygen} we plot $Q(\theta)$ in the region of interest that we found in our study of the triple-alpha reaction.

It is evident from the plot that incresing $\theta$ shifts the ground state of $^{16}O$ down, therefore the sub-threshold levels remain such. We see also that the level that could potentially give a resonance moves down by $\sim 120$ keV at the most, which is still 2.30 MeV above the threshold, still too far. We conclude that there are not any dramatic effects in the reaction $^{12}C + \alpha \to ^{16}O$, so that the ratio carbon/oxygen doesn't change appreciably, but even small values of the angle $\theta$ would result in a way greater abundance of both these elements.

\section{Conclusions}

The question raised in this paper can be phrased in the following way: would a non-zero angle $\theta$ change dramatically some aspects of nuclear physics? In order to find an answer, we singled out two examples, (i) the two-nucleon systems and (ii) the triple-alpha process, and studied the effects of $\theta$ on them. 

For (i) we found that the nuclear binding energies of deuteron and diproton would change by 10\% at $\theta \sim 130^\circ - 133^\circ$. Even if this effect does not look so dramatic, we believe that it would still affect the outcome of Big Bang Nucleosynthesis. For (ii) we found that, even for values of $\theta$ as small as $2^\circ$ or $3^\circ$, the abundance of carbon and oxygen would be ten times greater (see Figure \ref{fig:rratethetanarrow}) than what measured in our Universe. Would such a greater abundance still be consistent with the evolution of intelligent observer? We do not know with certainty the answer to this question. If negative, it would pose the anthropic bound that $\theta$ be less than $\sim 2^\circ$; if a factor of 1000 for the abundance, instead of 10, were not compatible with life, then the constraint on $\theta$ would be weaker: $\theta < 4.5^\circ$. 

We must stress that the numerical values given in (ii) are rough estimates. The main source of error is in the assumption that the energies of the excited states, with respect to the ground state, are not a function of $\theta$, which they most likely are, but it is very difficult to relate the spacing between these levels to first principles.

%We conclude that if $\theta$ were different from zero the consequences would not be catastrophic. Elements that are key for life, like carbon and oxygen would still be produced.

\section*{Acknowledgements}
I would like to thank Michael Dine for addressing me to this question and for his invaluable guide throughout this work.

\end{document}